\input{aipcheck}

\documentclass[
  draft            
  ]
  {aipproc}

\layoutstyle{8x11single}

\begin{document}

\title{A multi-scale approach to the physics of ion beam cancer therapy }



\author{A.V. Solov'yov}{address={Frankfurt Institute for Advanced Studies,
Ruth-Moufang-Str. 1, 60438 Frankfurt am Main, Germany}
}
\author{E. Surdutovich}{
  address={Frankfurt Institute for Advanced Studies,
Ruth-Moufang-Str. 1, 60438 Frankfurt am Main, Germany},altaddress={Department
of Physics, Oakland University, Rochester, Michigan 48309, USA}
}
\author{E. Scifoni}{
  address={Frankfurt Institute for Advanced Studies,
Ruth-Moufang-Str. 1, 60438 Frankfurt am Main, Germany}} 
\author{I. Mishustin}{
  address={Frankfurt Institute for Advanced Studies,
Ruth-Moufang-Str. 1, 60438 Frankfurt am Main, Germany}
  ,altaddress={Kurchatov Institute, Russian Research Center, 123182
  Moscow, Russia}}  
\author{W. Greiner}{
  address={Frankfurt Institute for Advanced Studies,
Ruth-Moufang-Str. 1, 60438 Frankfurt am Main, Germany}
}

\begin{abstract}
We propose a multi-scale approach to understanding physics related
to the ion/proton-beam cancer therapy and calculation of the
probability of the DNA damage as a result of irradiation of patients
with energetic (up to 430~MeV/u) ions. This approach is inclusive with
respect to different scales starting from the long scale defined by
the ion stopping followed by a smaller scale defined by secondary
electrons and radicals ending with the shortest scale defined by interactions of
secondaries with the DNA. We present calculations of the probabilities of
single and double strand breaks of the DNA and suggest a way of
further elaboration of such calculations.
\end{abstract}

\maketitle

\section{Introduction}

Ion-beam cancer therapies are being used more and more as
favorable alternatives to the conventional photon
therapy, also known as radiotherapy~\cite{Kraft05}\footnote{We limit
the references to either reviews or the most recent works that have caught our
attention and referred to other literature on the subject.}.  Their advantages
related to at the fundamental difference in the linear energy transfer (LET) by a massive projectile as compared with massless photons, namely by the Bragg peak
depth-dose distribution for the ion. It is due to this
peak, can the effect of irradiation on the tissue be more localized
increasing the efficiency of treatment and reducing the side effects. In order to
plan a treatment, a number of physical parameters, such as the energy
of projectiles, intensity of the beam, time of exposure,
{\em etc.}, ought to be defined. At present their definition is based on a set of
empirical data and experience of personal. Moreover, the optimization of treatment planning requires understanding of microscopic phenomena, which take place on time scales ranged from
$10^{-22}$~s to minutes or even longer times. Many of these processes
are not sufficiently studied. Thus, a
reconstruction of the whole scenario explaining, qualitatively and
quantitatively, the leading effects on each structural level scale presents a
formidable task not only for physics but also for chemistry, biology and medicine.  

\begin{figure}
  \includegraphics[height=.4\textheight]{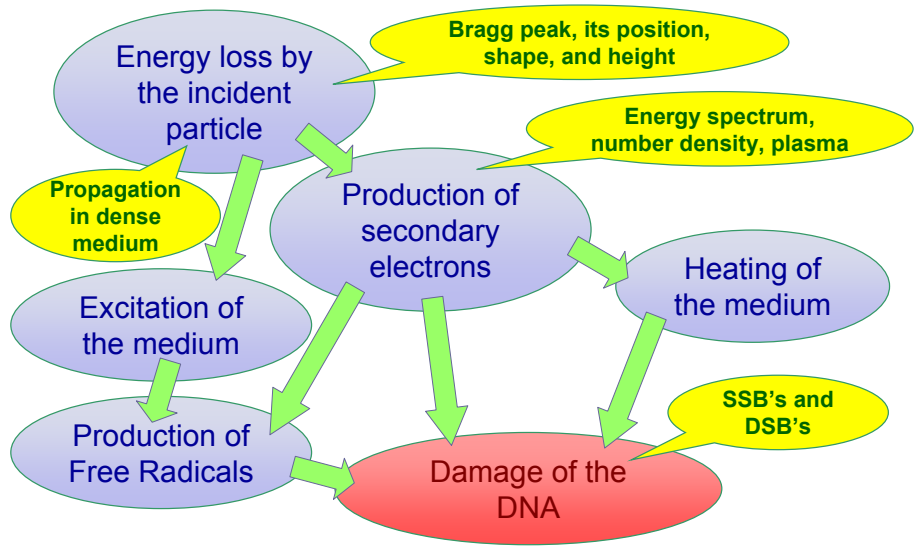}
 \caption{Schematics of the multi-scale approach.}
\label{scheme}
\end{figure}

The ultimate effect of the beam therapy is due to the DNA destruction
and subsequent killing of cells as a result of energy deposition by
the projectiles~\cite{Kraft05}.  Most of energy deposition by the ion
is due to ionization of a medium it traverse through, which by
about 75\% consists of liquid water. The secondary electrons formed in
this process are believed to be mostly responsible for the DNA damage,
either by directly breaking the DNA strands, or by reacting with water
molecules producing more secondary electrons and
of free radicals, which can also damage the DNA. One can also speculate about heating of the medium in the ion tracks making the DNA more vulnerable to damage, if not melting it. Among the
DNA damage types, we emphasize single strand breaks (SSB's) and double
strand breaks (DSB's). The latter ones are especially important because
they represent irreparable damage to the DNA. After a fast ion beam enters the tissue, many processes take place on different temporal and spatial scales until tumor cells die. The goal of our approach is to analyze these processes and select
the main physical effects which are responsible for the success of the ion-beam therapy. It turns out that several aspects play an important role as is illustrated in Fig.~\ref{scheme}.

The driving force of the ion-beam therapy is a propagation and stopping of the incident ions in
the tissue. Depending on the initial kinetic energy, the ions
penetrate to a certain depth and produce a Bragg
peak at the end of their range. Many works were devoted to this problem including both deterministic and Monte Carlo 
methods, see e.g. Ref.\cite{Pshen}. Using the information about cross sections of
atomic (such as ionization of water molecules) and nuclear (such as nuclear
fragmentation of projectiles) processes as an input, these models
give very good predictions of all characteristics of the Bragg peak,
its position, height, tail, etc. These
models provide a reasonable information on the energy deposition on a mesoscopic scale of about 0.1mm, which is sufficient for the treatment planning. On the other hand, this information is not sufficient to analyze the processes taking place on a microscopic
scale. The energy of ions changes from the initial energy in the range of 200-430~MeV/u down to about 50~keV/u. The next scale is defined by the secondary electrons, which are
produced as a result of ionization of molecules of the medium and by
radicals also formed as a result of energy loss by the projectile. The
maximum energy on this scale hardly exceeds 100~eV and the
displacement is of the order of 10-15~nm. So far, we have only considered ionization of water, a
surrogate for a biological medium, as a single process taking
place on this scale, leaving other atomic interactions for
later consideration. Ionization of the medium is the leading process
in accounting for the energy deposition by the ions we believe that the secondary electrons are mainly responsible for the DNA damage.
Interaction of electrons and radicals with DNA also happens on a nanometer
scale, and many works are devoted to study these
interactions~\cite{DNA2,DNA3,Sanche05,DNA1,DNA4,DNA5}.

 The main event of this scale is a diffusion of free electrons and radicals in the medium. Many chemical
reactions take place as well. They are also important for
estimates of the DNA damage since they define the agents interacting
with the DNA. Again, this aspect attracted plenty of attention of
Monte Carlo simulations adepts~\cite{Nikjoo06} who using various SDCS
for ion and electron energy loss (including the 
effects of the medium~\cite{liqh2o,Meesungnoen02,mfp,Pimblott3}) trace
the electrons and other 
species through the medium up to their interaction with the DNA. In
this paper, we present an approach to calculations that can be done on
this scale without using Monte Carlo simulations.

. The main input for our approach is the
singly differentiated cross section (SDCS) of 
ionization of water~\cite{epjd,nimb}. In this work, we use the
experimental results of Ref.~\cite{DNA2,DNA3,Sanche05}.

\section{Phenomena-defined scales}
\subsection{Ion stopping and production of secondary electrons}

The energy spectrum of secondary electrons was analyzed in
our previous works,~\cite{epjd,nimb}. It plays a major role in the energy
loss by projectiles and therefore determines to a large extent all characteristics of
the Bragg peak. As follows from Eq.~(\ref{mult}), it is also important for the DNA damage calculations (see below).

Physically, the SDCS is determined by the properties of the medium,
and since we use liquid water as a substitute for biological tissue,
it is determined by the properties of water molecule and the
properties of liquid water as continuous medium. All this information
is contained in real and imaginary parts of electric susceptibility of
liquid water. This approach can be generalized for any real tissue. In
order to do that, the same quantities, such as SDCS, of this medium
have to be used.  

In Refs.~\cite{epjd,nimb}, we used a semi-empirical parametrization by
Rudd~\cite{Rudd92} for the SDCS and obtained the position of the Bragg
peak with a less than 3\% discrepancy as compared to Monte Carlo simulations and
experimental data~\cite{epjd}. 
However, contrary to the calculations of total cross sections, or the Linear Energy Transfer (LET), where there is integration in $W$,  the calculations of the DNA damage may be more sensitive to the shape of the SDCS
at small energies, which for the liquid water is significantly
different from that of water vapor~\cite{Pimblott,Pimblott2,Dingfelder,Garcia}. 

The SDCS is a function of velocity of the projectile and, since
the ions are quite fast in the beginning of their trajectory, it has to
be treated relativistically. In Ref.~\cite{epjd}, this issue has been
solved by ``relativization'' of the Rudd parametrization by fitting it
to correct Bethe asymptotic behavior at the relativistic limit.   

Another important issue related to SDCS is the effect of charge
transfer that is due to picking-up electrons by the initially fully
stripped ions (such as $^{12}$C$^{6+}$) as they slow down in the medium. Since the SDCS is
proportional to the square of ion charge, its reduction
strongly influences such characteristics as the height of the
Bragg peak, secondary electron abundance, etc. In Ref.~\cite{epjd}, we
solved this problem by introducing an effective 
charge taken from~\cite{Barkas63}. As a result, the
effective charge of the $^{12}$C$^{6+}$ near the Bragg peak is about
$+3$ rather than $+6$.

Even after the relativistic treatment of the projectile and the
introduction of effective charge, the profile of the Bragg peak
obtained in our calculations was substantially higher and narrower
than those obtained by Monte Carlo simulations or experimentally. The
main reason 
for the discrepancy was that our calculations were performed for a
single unscattered ion, while in simulations as well as in experiments
the ultimate results are a combination of many ion tracks with a
significant spread in energy and position due to multiple scattering by water molecules. After we took
into account straggling of the ions, the shape of our Bragg peak
matched the shape predicted by the Monte Carlo simulations with nuclear fragmentation channels blocked (see for details in this collection the contribution of Ref.~\cite{RADAM2}). 

The nuclear fragmentation in the case of carbon
ions is quite substantial and should not be neglected. In principle,
we can include the beam attenuation due to nuclear reactions given the
energy dependent cross sections of these reactions. Then we would be
able to reproduce the attenuation of the ion beam, secondary electron
production due to different species, the spread of the Bragg peak due
to different penetration depths of different species, and the tail
following the Bragg peak due to light products such as protons and
neutrons. All these complications, however, were beyond our primary goal of
gathering most significant effects together and we leave it to the
future refining projects. We should mention that a successful treatment
of nuclear processes has been done by the Monte Carlo simulations~\cite{Pshen}.

Another process not accounted for in Ref.~\cite{epjd} is an excitation of
water molecules by the ions. This effect contributes in the energy
loss by the projectiles and shifts the Bragg peak towards the
source. Even though no secondary electrons are produced in this case, the
excitation channel may be important for the DNA damage since excited
water molecules may dissociate producing free radicals that, in their
turn, may damage the DNA. 

Thus, the ionization energy loss by the ions in liquid water is the dominating process for the ion
stopping and the energy spectrum of the secondary electrons.Additional energy losses are associated with excitation of water molecules leading to the production of free radicals. 
The SDCS defines both the shortest scale
related to the ion propagation and provides the initial conditions for the
next scale related to the secondaries. 

\subsection{Propagation of the secondary electrons}

Even though the SDCS that we have used in~\cite{epjd} may not be quite
adequate, this distribution gives some important predictions that
agree quite well with other calculations and measurements. Indeed, the
average energy of the secondary electrons,    
\begin{eqnarray}
\left\langle W \right\rangle= \frac{1}{\sigma_{\rm T}}\int_0^\infty W \frac{d\sigma (W,T)}{dW} dW~,
\label{wav}
\end{eqnarray}
 in the
vicinity of the Bragg peak ($T\approx 0.3$~MeV/u) is about 45~eV. This
value limits possible further scenarios for such
electrons. For instance, it was shown that such electrons
may excite or ionize another water molecule, but, most likely, only
once, and the next generation of secondary electrons is hardly capable
of ionizing water molecules~\cite{epjd,nimb}. This puts a cap on the amount
of produced secondary electrons.

The secondary electrons propagate
in the same medium as the ion, and interaction with the medium is
again determined by the SDCS with electrons being projectiles. The
interaction can be elastic or inelastic, and there is a probability of
stumbling on a DNA and causing its damage.

The angular distribution of the secondary electrons at energies about
and below 45~eV are rather flat~\cite{angle}. Therefore, to the first
approximation, we can
consider a Brownian motion of secondary 
electrons and use random walk or simply diffusion to describe their
propagation through the medium from a point on the ion trajectory
where they became unbound. The probability density to diffuse through the
distance $r$ after $k$ steps  is given by
the following~\cite{Chandra},
\begin{eqnarray}
R(k, r)=\frac{1}{\left(\frac{2\pi kl^2}{3}\right)^{3/2}}\exp\left(-\frac{3 
    r^2}{2 k l^2}\right)~.
\label{rwalk}
\end{eqnarray}
In this equation, the mean free path $l$ is the average distance that the
electron passes between two 
consecutive elastic collisions.The
elastic mean free path is determined by the SDCS of electrons and
we use the results of Ref.~\cite{mfp}. Its typical values are 0.3--1 nm for electron energies around 30 eV. The mean free path for inelastic collisions $l_{in}$ is typically many (about 10--20) times
longer. So, we assume that electrons mainly experience elastic collisions and
inelastic processes are included via attenuation factor $\exp\left(-l k/l_{in}\right)$, accounting a distance wandered $kl$ . 

Both  elastic and inelastic mean free paths depend on the energy of the 
wandering electron. This energy is changing gradually, and strictly speaking, the mean free path is a
function of the initial energy and the number of steps. However, the
estimates of ref.\cite{Meesungnoen02} suggest that the 
electron energy loss over typical distances of 6-10~nm  is not very large.
Therefore, we will use some average energies and
assume that they do not change during the diffusion.

\subsection{Evaluation of the DNA damage}
\label{ionpass}

\begin{figure}
  \includegraphics[height=.4\textheight]{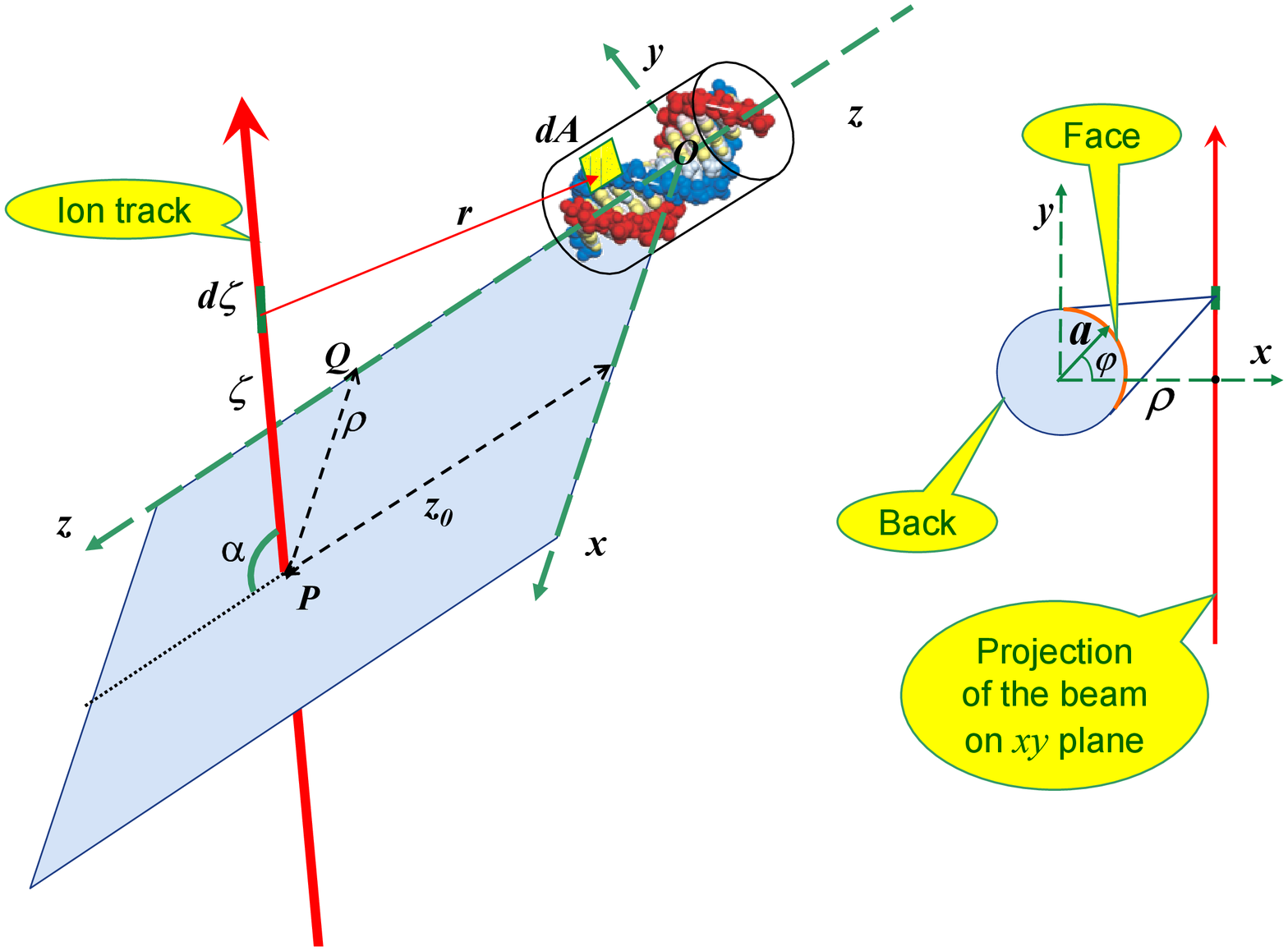}
  \caption{Geometry of the model: $z$ is the cylindrical axis of the DNA convolution and  $x$ is chosen to be parallel to $PQ$, the line of closest approach between $z$ and the beam (ortogonal to both), of length $\rho$, at distance $z_0$ from the center of the convolution $O$. $\zeta$ is the coordinate of any point in the beam with respect to $P$ and $\alpha $ is the angle between the beam and $z$. In the right panel we show a projection on the $xy$ plane, where $a=1.1$nm is the DNA radius and $\phi$ is the polar angle defining a point on its surface.}
\label{geom}
\end{figure}

A DNA damage, such as a SSB, is a result of a sequence
of mutually independent events. First, secondary electrons with a certain kinetic energy $W$ are produced at a certain depth $x$. Later on
they interact with a surrounding medium via elastic and
inelastic collisions, and gradually lose energy until become bound. Depending on the electron's energy, momentum and position, there is a chance that an
electron stumbles on a DNA molecule and damages it.
Fig.~\ref{geom}. 
To evaluate the DNA damage by secondary electrons we propose the following model. First, we represent the DNA molecule by a single convolution, i.e. a cylinder of length 3.4~nm and radius 1.1~nm (these parameters are well estabilished experimentally). The reason for that is that the DSB's are defined as simultaneous breaks of both DNA strands located within the single convolution. Second we assume that this cylinder is irradiated by the flux of secondary electrons produced by the ion traversing the medium at a certain distance from the DNA. This geometrical picture is schematically shown in Fig.~\ref{geom}. It is clear that the mutual location and orientation of the DNA cylinders and ion trajectories are randomly distributed. Therefore, the total damage to the DNA molecules can be calculated by averaging over all the possible configurations.
From that scheme stems that any distance $r$ between a point on the track and a point on the cylinder is given by
\begin{eqnarray}
r^2=(a\cos \varphi - \rho)^2+(a\sin\varphi - \zeta \sin
\alpha)^2+(z-z_0-\zeta \cos \alpha)^2
\label{r2}
\end{eqnarray}

The flux of secondary electrons through a unit area at a distance $r$ from the production point is 
 \begin{eqnarray}
 {\vec \Phi}_k(r, W, W_f)=D\nabla R(k, r)\frac{d^2 N}{dW d\zeta}(\zeta, W)\Delta W~,
\label{flux}
\end{eqnarray}  
where $D=kl^2/6$ is the diffusion coefficient multiplied by the
average time of wandering.   The last factor in Eq.~\ref{flux} is the number of electrons with energies between $W$ and $W+\Delta W$ produced from the segment of ion trajectory $d\zeta$. It can be obtained from the singly differentiated ionization cross section (SDCS), as explained in refs.~\cite{epjd,nimb}.
$W_f$ is the electron's energy when it reaches the surface after traveling from  the track.
Finally, we assume that the number of the SSB's within the DNA cylinder is proportional to the number of electrons crossing the surface,
\begin{eqnarray}
N_{SSB}=\Gamma_{SSB}(W_f)\sum_k \int d\zeta d{\vec A}\cdot {\vec \Phi}_k(r, W, W_f)~,
\label{mult}
\end{eqnarray}
where the integration is done over the surface of cylinder and the ion trajectory. The unknown quantity $\Gamma_{SSB}(W_f)$, that for the moment we assume as a constant, should be determined from experimental data. 
The calculation for a general case as shown in Fig.~\ref{geom} is rather cumbersome.
Therefore, let us set $z_0$ to zero,
and consider separately two limiting cases, the ``parallel'' case when
$\alpha=0$, and the ``normal'' case when $\alpha=\pi/2$. 
In the parallel case, the cylinder containing the DNA convolution is
parallel to the ion track and $\rho$ is the distance between the axis
of the cylinder and the track. In the normal case, the axis of this
cylinder is perpendicular to the ion track and when $z_0=0$, $\rho$ is again
the distance between the axis of the cylinder and the track and the
beam projects along $\rho$ to the center of the cylinder. 
In both of these cases we need to set the limits for angular
integration over $\varphi$. 

Looking from any point of the ion track, there are two surfaces of the
DNA cylinder: the ``front'' or ``face'' surface and the ``back''
surface (see Fig.~\ref{geom}). In our model, if a wandering electron
hits the face or the back surface, it may cause a strand break with a certain
probability. Therefore we
simply add the probability of a SSB due to electrons striking the back
surface to that for the face surface regardless of directions of their
motion leaving an introduction of attenuation mechanism accounting for
electron passage ``through the DNA'' to an extension of this model. 

\begin{figure}
  \includegraphics[height=.3\textheight]{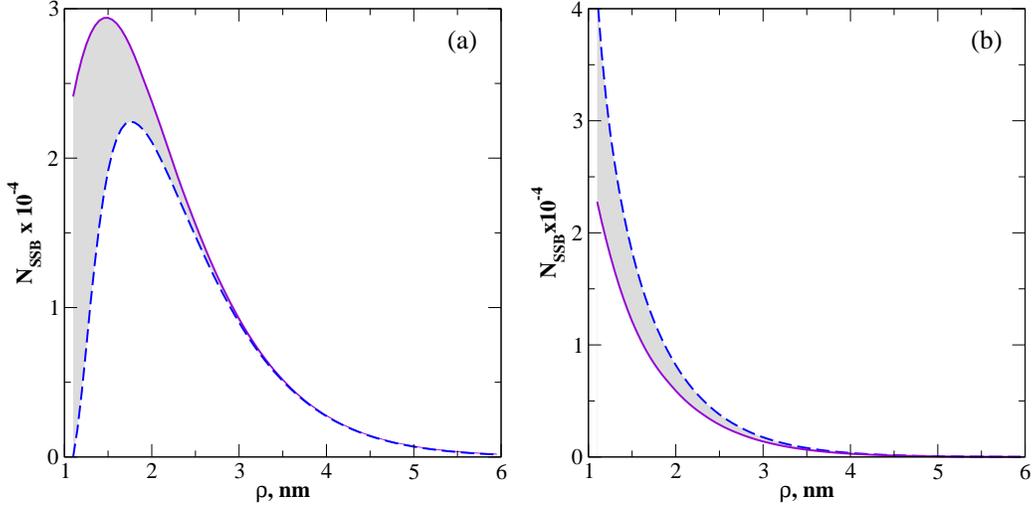}
  \caption{A comparison of numbers of SSB's for parallel (dashed line)
  and normal 
  (solid line) configurations on the face (a) and on the back side (b)
  for 20-eV electrons.}
\label{parandnorm}
\end{figure}

\begin{figure}
  \includegraphics[height=.3\textheight]{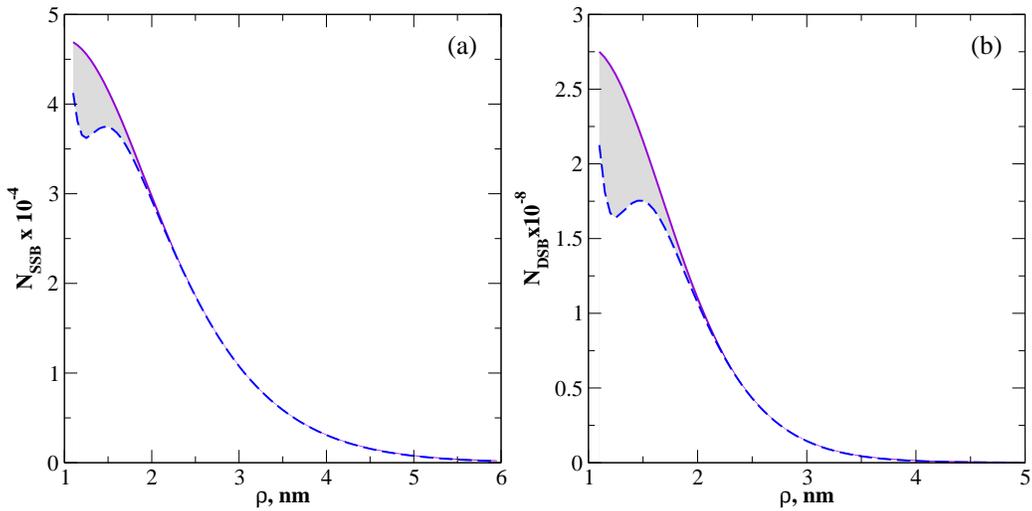}
  \caption{A comparison of dependencies of overall (due to the whole
  surface of the cylinder) SSB's (a) and
  DSB's due to separate 
  electrons (b) on distances to the DNA convolution in the parallel
  (dashed line)  and normal (solid line) cases for 20-eV electrons.}
\label{ssbdsb}
\end{figure}

\begin{figure}
  \includegraphics[height=.3\textheight]{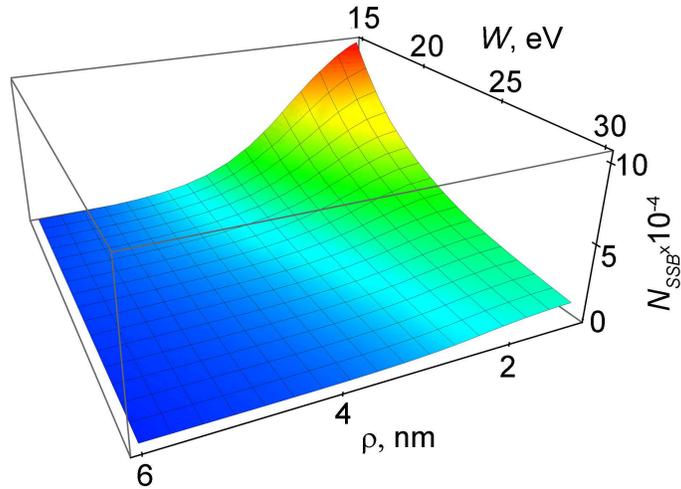}
  \caption{The number of SSB's dependent on the distance to the DNA
  convolution,
  $\rho$, and the energy of secondary electrons (parallel case).}
\label{ssbfig}
\end{figure}

The results of the integration are shown in Fig.~\ref{parandnorm}. 
All
curves show the dependence on the distance $\rho$ from the DNA to the
beam. When the distance is large enough ($>3$~nm) the normal and
parallel cases coincide for the face side as well as for the back
side, only at small distances there is a difference. This difference
is significant only for back and face sides taken separately, but not
for their combination shown in Fig.~\ref{ssbdsb}a. 
This means that the
geometrical details of orientation of DNA segments with respect to the
beam may not be so significant, since all varieties lie somewhere
in the shadowed region in Fig.~\ref{ssbdsb}a between the two
curves. For an analysis of a more general picture, some average curve
(lying between the two curves in Fig.~\ref{ssbdsb}) should be used
with $\rho^2$ replaced by $\rho^2+z_0^2$.   
 In this calculations we took the value  $\Gamma_{SSB}(W_f)=5\times 10^{-4}$, extracted from experimental data of Refs.~\cite{DNA2,DNA3,Sanche05}, where SSB's and DSB's were induced by the electron beam of energy 0.1--30 eV. The density of the beam was such that only a small fraction of DNA molecules was irradiated.
The numbers of SSB's caused by the secondary electrons
depending on the distance $\rho$ and 
the energy of the secondary electrons for the parallel case are shown in
Fig.~\ref{ssbfig}. 
The decline of the number of SSB's with increasing $\rho$ is an explicit
consequence of Eq.~(\ref{rwalk}) with account of (\ref{r2}). The
energy dependence is mainly caused by the dependence of mean free
paths on the energy and the attenuation factor given
above. This factor is heuristic and may have to be
corrected later when the corresponding experimental or computational
data are available.

What about the double strand breaks? From Ref.\cite{DNA2,DNA3,Sanche05}, it
follows that the DSB's produced by the electrons with energies over
about 5~eV are caused in one hit, {\em i.e.}, if a particular electron
with a probability of about 0.0005 causes a SSB, the same electron
causes a DSB with a probability of about 0.0001. This is why the
analysis of the probabilities of SSB's are so important. If the energy
of the secondaries are high enough they give the probability of the
DSB's after being divided by some factor of about 5. Therefore,
Fig.~\ref{ssbfig} also gives the shape of the main contribution to the
DSB's. This, of course, strongly depend on the electron density and energy. Many authors e.g. \cite{mfp} use a criterion that electrons with energy above 
a certain threshold (say 30 eV) are needed to produce a DSB.
At energies
lower than 5~eV the situation changes: one electron does not
cause two breaks. Therefore, we need to calculate the number of DSB's
caused by two different electrons. From the geometry of a DNA convolution
we infer that the probability of a DSB is proportional to
$\frac{1}{4}(N_{SSB, face}^2+N_{SSB, back}^2)+\frac{1}{2}N_{SSB,
face}N_{SSB, back}$.  
The numbers for the DSB's caused by different electrons in parallel
and normal cases are shown in Fig.~\ref{ssbdsb}b. As well as in the
case of SSB's the difference due to geometry is not very significant.
Even though the numbers of DSB's are many times smaller than those in
Fig.~\ref{ssbdsb}a, this effect may be significant if the density of
secondary electrons is large enough. According to our estimates in
Refs.~\cite{epjd,nimb}, the density of the secondary electrons in the
vicinity of the Bragg peak at therapeutic conditions is by about 16
orders of magnitude higher than the electron density in experiments of 
Ref.~\cite{DNA2,DNA3,Sanche05}. Therefore, this effect may be an
important correction to the paradigm.

This concludes our approach to calculations of DSB's and SSB's due to
secondary electrons produced by the ions. What about secondaries
produced by the electrons and free radicals produced by the ions? Can
they be treated the same way as we have treated the secondary
electrons in the previous subsections?

\subsection{Other secondaries}

Secondaries that can be treated almost in the same way as the
secondary electrons that we just went through are OH$\cdot$
radicals. They are formed as a result of ionization of water molecules
by the ion after dissociation of water ion into OH$\cdot $ and
H$^+$. These radicals are formed almost at the same place as the
secondary electrons. The difference is, of course, in a different
diffusion coefficient, and different time of getting to the
DNA, which is by about 100 times longer than that for secondary
electrons. Then, the DNA damage caused by OH$\cdot$ may also be 
different~\cite{sonntag,geront}. Nonetheless, if the effect produced by
OH$\cdot$ is an important player, this is a recipe of its inclusion. 
The same can be said about those free radicals that are formed as a
result of excitation of water molecules by the ions. These radicals
(OH$\cdot$ and H$\cdot$) are also produced on the ion trajectory and can
be treated in the same way as secondary electrons.

The other secondaries, such as the second generation of electrons
produced by the first generation via ionization of water, radicals
produced as a result of this process and the radicals  H$\cdot$
produced via interaction of secondary electrons with water molecules
({\em e.g.}, through dissociative attachment) can be treated in the
following way. Let the interaction that produces a ``desired agent''
happens at some point ${\vec r'}$. Then the previous procedure has to
be divided into three parts: diffusion of the secondary electron from
the point of origin (the ion's trajectory) to ${\vec r'}$, an
interaction that leads to the production of the agent at ${\vec r'}$,
and the diffusion of the agent to the DNA cylinder. Then, perhaps
cumbersome, integration over ${\vec r'}$ has to be performed. 

\section{conclusions}

Thus we presented a multi-scale inclusive approach to the physics
relevant to the ion-beam cancer therapy. We intend to present a clear
physical picture of 
the events starting from an ion entering a tissue leading to the DNA
damage as a result of this incidence. We view this scenario as a palette of
different phenomena that happen on different time, energy, and distance
scales. From this palette, we choose major effects that adequately
describe the leading scenario and then describe the ways of inclusion
of more details. We think that calculations in this field can be
made inclusively without dwelling on a particular scale. Our
calculations are time effective and 
can provide a desired accuracy. They show that the seemingly
insurmountable complexity of geometry of the DNA in different states
may be tackled because the geometrical differences, shown in Fig.~\ref{ssbdsb},
are insignificant. We would like to encourage the experimentalists to
provide data more relevant to the actual conditions of irradiation,
especially on the smallest scales involving the DNA damage. This
information is vital for further tuning of our approach by selecting
and elaboration on the most important aspects of the scenario.

\begin{theacknowledgments}
This work is partially supported by the European Commission within the
Network of Excellence project EXCELL and the
Deutsche Forschungsgemeinschaft; E.S. is grateful to A.V. Korol and
I.A. Solov'yov for multiple fruitful discussions and to FIAS for
hospitality and support.  
\end{theacknowledgments}

\end{document}